\documentclass[twocolumn, a4paper]{article}

\usepackage[pdftex]{graphicx}
\newcommand{\OPF}{Openpipeflow }

\renewcommand{\vec}[1]{\mbox{\boldmath $#1$}}

\usepackage[usenames,dvipsnames]{xcolor}
\newcommand{\rev}[1]{#1} %{{\color{blue}#1}}

\usepackage{amssymb}
\begin{document}

%\begin{frontmatter}



\title{The Openpipeflow Navier--Stokes Solver}


\author{Ashley P.\ Willis %}
%
%\address{
\\  {\it
School of Mathematics and Statistics, 
University of Sheffield,} \\ {\it South Yorkshire, S3 7RH, U.K.}
}
\maketitle

\begin{abstract}
%Ca. 100 words
%Pipe flow is a simple and familiar set up, yet the 
%flow patterns that emerge exhibit rich chaotic dynamics. This provides a setting for investigating the principles of simulation at one level, and at another, for developing new methods designed to probe fundamental properties of dynamical systems.
%
Pipelines are used in a huge range of industrial processes involving 
fluids, and 
the ability to accurately predict properties of the 
flow through a pipe is of 
fundamental engineering importance.  
%Applications range from huge international oil pipelines,
%to water supply and heating systems in the home.
%
Armed with parallel MPI, Arnoldi and Newton--Krylov solvers, 
%(designed to integrable with any simulation code) 
%and parallel MPI, 
the \OPF code can be
used in a range of settings, from large-scale simulation of highly 
turbulent flow, 
to the detailed analysis of nonlinear invariant solutions (equilibria and periodic orbits) and their influence on the dynamics of 
the flow.
\\[10pt]
{\bf Website:}~~\texttt{openpipeflow.org}\\[2pt]
{\bf Reference:}~~SoftwareX, 6, 124-127.\\[2pt]
{\bf DOI:}~~10.1016/j.softx.2017.05.003

\end{abstract}

%\begin{keyword}
%CFD 1 \sep Dynamical systems 2 %\sep Newton--Krylov 3



%\end{keyword}

%\end{frontmatter}

%\linenumbers


%Description of your software in maximum 6 pages.

\section{Motivation and significance}
\label{}

%Introduce the scientific background and the motivation for developing the software.

%Explain why the software is important, and describe the exact (scientific) problem(s) it solves.

%Indicate in what way the software has contributed (or how it will contribute in the future) to the process of scientific discovery; if available, this is to be supported by citing a research paper using the software.

%Provide a description of the experimental setting (how does the user use the software?).

%Introduce related work in literature (cite or list algorithms used, other software etc.).

The flow of fluid through a straight
pipe of circular cross-section is a canonical
setting for the study of stability, transition and properties 
of turbulent flow. At low 
flow rates, the 
flow everywhere is in the direction parallel to the axis of the pipe, 
a simple `laminar' flow.  At larger flow rates it typically
undergoes a transition to a complex `turbulent' flow, characterised by 
an abundance of swirling eddies.
As early as 1883, Reynolds observed that the transition from 
laminar to turbulent flow is highly dependent on perturbations of finite
amplitude to the initial flow
\cite{R1883}.\footnote{Reynolds referred to what we now call `laminar' and `turbulent' flows by `direct' and `sinuous' flow, respectively.}
Nevertheless, he also noticed that the appearance of turbulence is consistent with respect to the value of the non-dimensional combination $D\,U/\nu$, at around 2000,
where $U$ is the mean axial speed, $D$ the 
diameter of the pipe, and $\nu$ the kinematic viscosity.
This combination is the now 
famous Reynolds Number, $Re=D\,U/\nu$, 
used in a huge range of systems involving fluids, 
where $D$ and $U$ are typical length and velocity scales for the system.

It has been known for some time that the Navier--Stokes equations
together with the no-slip boundary conditions
 accurately predict the evolution of the flow pattern, e.g.\ the landmark prediction of supercritical transition to a roll pattern
for the flow of water between rotating cylinders
by G.\ I.\ Taylor \cite{Taylor23}
(transition due to linear instability beyond a critical rotation
rate). Despite this development and the legacy of the work of Reynolds, the nature of subcritical transition (transition in the absence a linear instability)
and the dynamics of pipe flow has largely remained a mystery. But much
has changed following the discovery finite-amplitude solutions to the 
Navier--Stokes equations, for pipe flow as recently as 2003 \cite{FE03}.
These solutions, often referred to as `exact coherent states' \cite{W01}
are believed to embody the processes that sustain 
turbulence to and form a `skeleton' for the dynamic paths taken by the 
evolving flow patterns.
Comprehension of the nonlinear dynamics, 
particularly of transition in pipes, and likewise in Couette and channel flows, has progressed in leaps and bounds over the last decade, based on the study of these solutions.  New more general families of solutions  
continue to be discovered, and their unstable manifolds are just 
beginning to be calculated
\cite{Pringle09,deLMeAvHo12,AvMeRoHo13,ChWiKe14}.

The code that has evolved into \OPF  
has played a significant role
in the realisation of this odyssey. 
\rev{\OPF offers a more simplified approach than large computational
fluid dynamics (CFD) packages -- the aim during development has been 
to maintain a compact and readable code.  
Thus \OPF is easily adapted 
for a given analysis and extendible to new numerical methods.}
The code has recently been upgraded with a substantially improved parallelisation, and continues to
be augmented with new %utilities %for new methods and 
extensions, \rev{for example large-eddy simulation (LES)}.

Following the rapid expansion of computational resources that has occurred
in recent times, 
pipe flow is a prime example of a `high-dimensional' system
that is receiving examination with methods previously limited to systems
with only a few degrees of freedom, such as the Lorenz attractor or the
Kuramoto--Sivashinsky equation; see e.g.\ \cite{AMdABH11,WiShCv15}.
In the other direction, observations from large-scale simulations of pipe 
flow have inspired low-order
models \cite{barkley2015rise,ShHsGo15}.
 Pipe flow also provides a simple setting for the development
of computationally intensive new methods, such as adjoint optimisation techniques, e.g.\ \cite{pringle2012minimal}.

\section{Software description}
\label{}

%Describe the software in as much as is necessary to establish a vocabulary needed to explain its impact. 
\OPF implements a second-order predictor-corrector scheme, with
%optional 
automatic time-step control, for simulation of 
flow on the cylindrical domain 
$(r,\theta,z)\,\in\,[0,1]\times[0,2\pi/m_p)\times[0,2\pi/\alpha)$,
where $m_p$ and $\alpha$ are parameters that determine spatial
periodicity.
Variables in the Navier--Stokes equations are discretised in the form
\begin{equation}
   A(r,\theta,z)\,=\,\sum_{k<|K|}\,\sum_{m<|M|} \rev{A_{km}(r_n)} \, 
	 \mathrm{e}^{\mathrm{i}(\alpha k z + m_p m \theta)} \, , 
   %\quad n=1..N\, ,
\end{equation}
$n=1..N$, 
\rev{
where the points $r_n$ are %a set of points %, $n=1..N$.  
%These points %$r_n$, $n=1..N$ 
%may be arbitrarily 
distributed on $[0,1]$.
By default the $r_n$ are located at the roots of a Chebyshev polynomial, bunched towards the boundaries to resolve large gradients that occur in the boundary layer.\footnote{\rev{Optionally the $r_n$ may be read in from a file,
\texttt{mesh.in}.  In LES simulations, for example, it may be desirable to 
specify the distribution of points with respect to the position of the 
turbulent buffer layer.}}
Derivatives in the radial dimension are calculated using finite differences, so that they may be evaluated using banded matrices.
The number of points used, and hence the width of the bands, 
is an integer parameter; by default derivatives are calculated using 9 points, for which 1st/2nd order derivatives are calculated to 8th/7th order.}
Following the $3/2$ 
dealiasing rule, the
sums are evaluated on $3K\times3M$ grids in $z$ and $\theta$ respectively.
Periodicity in $z$ is a commonplace approximation 
\rev{that has been shown
to capture all the relevant physics of turbulent flow \cite{Eggels94}
and the transition to turbulent flow \cite{AMdABH11}}.
The dimension $\theta$ is naturally periodic ($m_p=1$).
Rotational symmetry ($m_p=2,3,...$) is often applied,
\rev{since finite-amplitude solutions typically
satisfy rotational symmetry, or applied simply to reduce computational expense when
the structures of interest are much smaller than the domain,
e.g.\ near-wall vortices at large flow rates.}
%The set of finite-difference points
%$r_n$, $n=1..N$  may be arbitrarily distributed on 
%$[0,1]$, but by default
%are located at the roots of a Chebyshev polynomial, bunched towards the
%boundaries to resolve large gradients that occur in the boundary layer.
%where large gradients occur. 

A pressure-Poisson equation (PPE) formulation is employed
and an influence-matrix technique applied for the enforcement of 
boundary conditions
 %Application
%of an influence-matrix technique to this formulation
%, the mathematics for which may be found in 
\cite{guseva2015transition}.
%, where a sister-code for Taylor--Couette flow is described.
Let $\vec{g}$ be 
a vector of boundary conditions, written such that $\vec{g}=\vec{0}$ when
they are satisfied. 
The influence-matrix technique has several nice features.: 
\begin{itemize}
\item Alternative boundary conditions, e.g.\ slip or oscillations, are easily 
introduced by changing the single function that evaluates $\vec{g}$;
%\item It is easy to recover the true pressure. In many methods that use 
%alternative boundary conditions, e.g.\ the popular fractional-step projection
%approach, the projection function does not usually correspond to the
%pressure $p$, and is often denoted $\Pi$ to make the distinction.
\item %The resulting error 
The usual no-slip and divergence conditions at the boundary are satisfied
%in all boundary conditions, i.e.\ $\|\vec{g}\|$ 
%is 
such that $\|\vec{g}\|$ is typically at 
the level of the machine epsilon for the given floating-point precision;
\item Computational overhead is negligible compared to evaluation of 
non-linear terms;
\item No stability issues have been observed.
\end{itemize}

Utilities and templates for runtime- and post-processing are 
included, including a Newton--Raphson solver for the calculation and 
continuation of invariant solutions. 
The Newton solver for the pipe flow, which has
a multiple-shooting option (orbits may be split into multiple sections), calls
a utility that implements a combined Krylov--Trust-region approach 
\cite{Visw07b}.
This Newton--Krylov--Trust-region utility is designed to be integrable with
any simulation code.

\OPF  
is written in Fortran90 and uses basic modules and derived types. 
Esoteric extensions to the programming language have been deliberately avoided. The code makes use of FFTW, LAPACK and NetCDF libraries. 
Optionally, for parallel use an MPI library is required.

\begin{figure}[h]%
\includegraphics[width=\columnwidth]{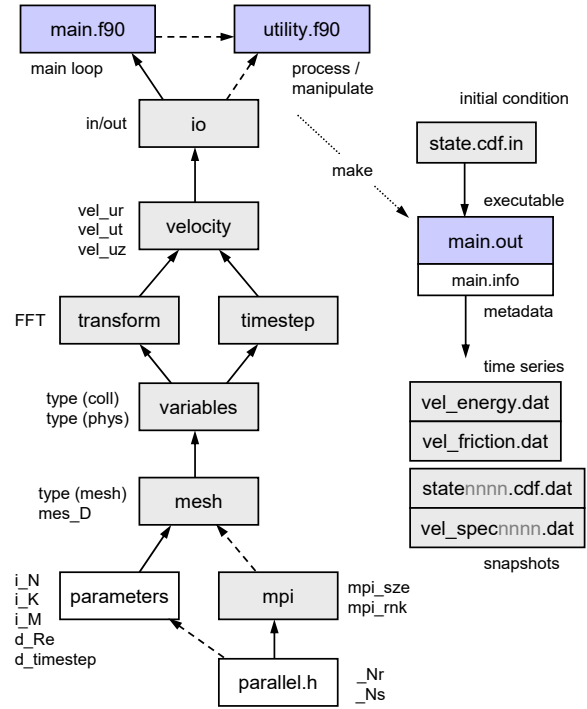}%
\caption{Code structure and program interaction. 
The MPI library is not required if
$\texttt{\_Nr}=\texttt{\_Ns}=1$. 
To post-process data it is sufficient for a utility to inherit the io module. To
process at run time, it is possible to inherit the whole main loop.
}%
\label{fig:modules}%
\end{figure}

\subsection{Software Architecture  and Functionality}
\label{}
%Give a short overview of the overall software architecture; provide a pictorial component overview or similar (if possible). If necessary provide implementation details.

%\subsection{Software Functionalities}
%\label{}
%Present the major functionalities of the software.

%\subsection{Sample code snippets analysis (optional)}
%\label{}

See Fig.\ \ref{fig:modules} for a schematic of the code structure and program 
interaction.  Once parameters are set and the code built, most jobs begin with a single initial condition, \texttt{state.cdf.in}. Outputs from another job, \texttt{statennnn.cdf.dat}, usually make the best initial conditions (\texttt{nnnn} is a 4-digit numeric label). A variety of possible initial conditions are provided in the database at openpipeflow.org. Truncation or interpolation of initial conditions with a different resolution is automatic.

A selection of utilities, plus templates for post-processing or 
runtime-processing, are described in the online manual.

\subsection{Implementation details}

Linear systems that originate from the implicit solution of the 
viscous terms in the Navier--Stokes equations
% for independent Fourier modes 
are solved using banded
matrices and LU-decomposition for each Fourier mode.  Nonlinear
terms are evaluated pseudospectrally.
%By default seven points are involved in the finite-difference stencils.
% equiv to spectral collocation method if all points used.

Parallelisation is achieved via a split into \texttt{\_Nr} radial and 
\texttt{\_Ns} axial sections,
\rev{and the work is divided over
$\texttt{\_Np} = \texttt{\_Nr}\times\texttt{\_Ns}$ cores}
 (\#-defined symbols in \texttt{parallel.h}). 
Due to the form of the data transposes
involved in the transforms between `collocated' (Fourier) and physical space
(\texttt{type (coll)} and \texttt{type (phys)}), 
the number of cores is limited to $N\times M$. 
This has been a distant limitation to date.

The recent upgrade to the two-dimensional split from the  
one-dimensional `wall-normal' split (independent 2D-FFTs) not only extends the
maximum number of cores from $N$ to $N\times M$, but also reduces the number of messages that must be sent. The transform involves two stages of
FFTs and transposes, but each transpose involves only \texttt{\_Nr} or 
\texttt{\_Ns} cores.  
\rev{For a transpose involving $p$ cores, each core must send $p-1$ messages.}
Therefore, choosing $\texttt{\_Nr}\approx\texttt{\_Ns}
\rev{~\approx\sqrt{\texttt{\_Np}}}$, 
the number of messages is $O(2\sqrt{\texttt{\_Np}})$ versus 
$O(\texttt{\_Np})$.
This can substantially reduce time lost in latency due to the time setting up communications. \rev{Further details can be found on the Core Implementation page of the 
online manual.}

\section{Illustrative Examples}
\label{}

%Provide at least one illustrative example to demonstrate the major functions.

%Optional: you may include one explanatory video that will appear next to your article, in the right hand side panel. (Please upload any video as a single supplementary file with your article. Only one MP4 formatted, with 50MB maximum size, video is possible per article. Recommended video dimensions are 640 x 480 at a maximum of 30 frames/second. Prior to submission please test and validate your .mp4 file at $ http://elsevier-apps.sciverse.com/GadgetVideoPodcastPlayerWeb/verification$. This tool will display your video exactly in the same way as it will appear on ScienceDirect.).

\subsection{Modelling a Coriolis force}

Does the Coriolis force, an extra force term due to rotation of Earth,
affect the flow in experiments?
\rev{The file \texttt{utils/Coriolis.f90} is an 
example utility is provided with the distribution that models this case.}

The main loop of the core code \rev{
already includes several calls to a null function at key points during
the timestepping process; see  
\texttt{var\_null(flag)} in \texttt{program/main.f90}.
The \texttt{flag} may be used to detect the stage at which the
function has been called.
Here, we replace the null function with the function in \texttt{Coriolis.f90}
% in our 
%utility \rev{\texttt{utils/Coriolis.f90}}.
%\begin{verbatim}
%#define var_null coriolis
%#include main.f90
%subroutine coriolis(flag)
%   integer, intent(in) :: flag
%   if(flag/=2) return
%   ...
%\end{verbatim}
and detect the case \texttt{flag==2},} which
indicates that nonlinear terms have just been
evaluated. 
At this point we add the Coriolis forces to the nonlinear terms. % at this stage.  
Note that no changes to the core files,
including \texttt{main.f90}, are necessary.

\begin{figure}%
\begin{center}
\includegraphics[width=0.7\columnwidth]{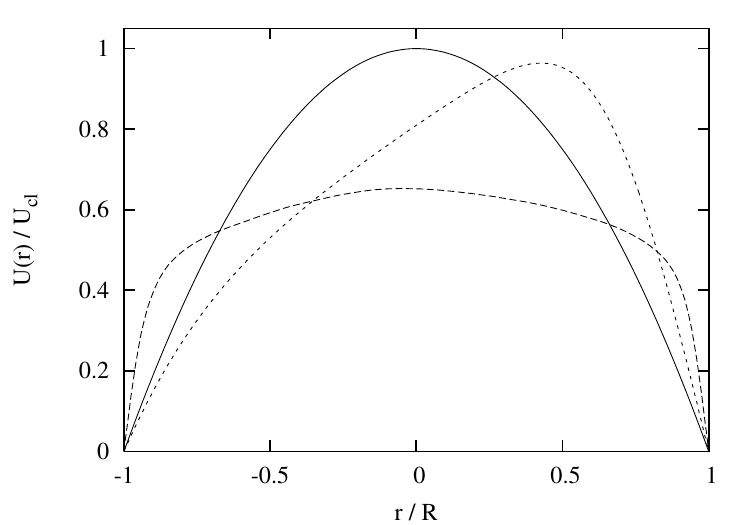}%
\end{center}
\caption{Response of flow at $Re=5300$ to a Coriolis force. 
%U$_\mathrm{cl}$ is the centreline speed
%for laminar flow at this mean flow rate, and $\mathrm{R}=D/2$ is the pipe radius. 
Solid: Laminar flow, $E\to\infty$ (no rotation).  
Short-dash: Laminar flow, $E=1$.
Long-dash: Turbulent flow, $E=1$.
}%
\label{fig:Coriolis}%
\end{figure}

Figure \ref{fig:Coriolis} %lagrange:~/run/Eggles2/7900
shows the mean axial flow profile for 
laminar and turbulent 
flow at an Ekman number $E = \nu / (2\Omega D^2)=1$
for a pipe with axis oriented east-west, 
\rev{perpendicular to the rotation of axis for any latitude.}
%The rotation axis is chosen to point in the Cartesian $x$-direction
For a pipe %oriented east-west and 
filled with water at 20$\,^\circ$C,
this corresponds to a diameter $D$ of approximately 8.3cm;  
$Re = 5300$ in all cases,
U$_\mathrm{cl}$ is the centreline speed
for laminar flow at the same mean flow rate, and $\mathrm{R}=D/2$ is the pipe radius. 
For this $Re$, laminar 
flow shows a substantial response, \rev{and the profile is similar to
those reported in \cite{draad1998earth}}.
Turbulent flow, however, shows no asymmetry.  
The turbulent mean profile is indiscernible from the 
documented test case \cite{Eggels94}.

\subsection{Unstable manifold of a travelling wave solution}

\begin{figure}%
\includegraphics[width=\columnwidth]{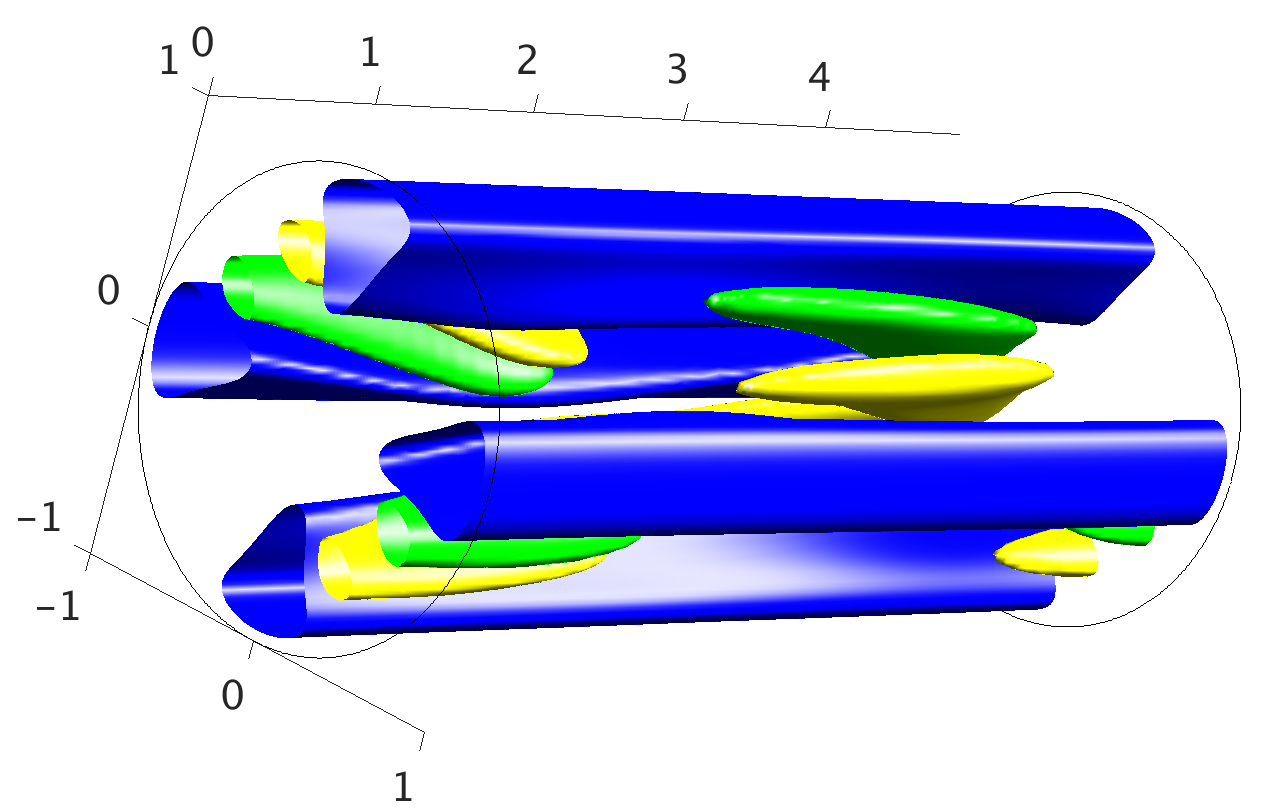}%
\caption{N2\_ML solution. 
(blue) Slow `streaks' -- axial flow slower than the 
mean flow profile by $>0.07$.
(yellow and green) `vortices' -- axial vorticity $>0.2$ and $<-0.2$.}%
\label{fig:TWvis}%
\end{figure}

\begin{figure}%
\includegraphics[width=\columnwidth]{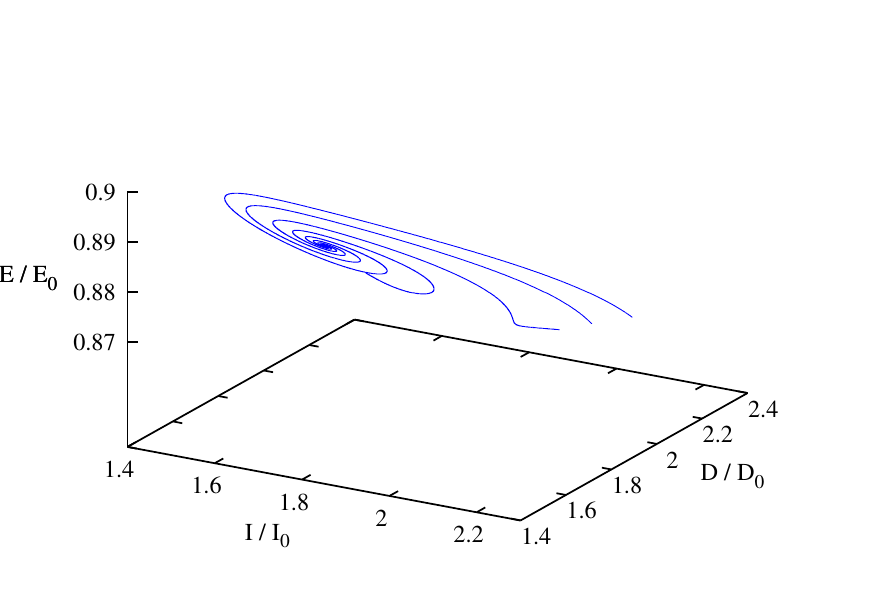}%
\caption{Projection of the unstable manifold of the 
N2\_ML travelling wave solution.}%
\label{fig:spiral}%
\end{figure}

A travelling wave solution is an equilibrium when considered in a frame
moving at its phase speed. 
In this case we consider
the `upper branch' solution known as N2\_ML, Fig.\ \ref{fig:TWvis},
which in its symmetry subclass has a single unstable complex
eigenvalue, 
\rev{$0.00620+0.0183\,\mathrm{i}\,(\mathrm{U}_\mathrm{cl}/\mathrm{R})$ 
(after one rotation it expands by a factor $8.4$)};
$Re=2400$, $\alpha=1.25$, $m_p=2$; see \cite{ACHKW11} for 
further details. 
For a given nearby state, the Newton--Krylov utility 
(\texttt{newton.f90})
can find such solutions and output their leading unstable eigenvectors
\rev{(solution \texttt{state1000.cdf.dat} and real and imaginary parts
of the leading eigenvector, 
\texttt{state1001.cdf.dat}, \texttt{state1002.cdf.dat};
available at the online Database).}
To visualise the unstable manifold, we use a utility (\texttt{addstates.f90})
to add small multiples \rev{($\approx10^{-4}\times$)} of the real part of the eigenvector to the solution, then use these as initial conditions 
(\texttt{state.cdf.in}) for a set of simulations.
Figure \ref{fig:spiral} %lagrange:~/run/Predrag2/7900
shows a projection of the unstable
manifold of N2\_ML, as an outward spiral, with deformation at larger amplitudes
due to nonlinearity.  The coordinates are the 
kinetic energy $E$, energy input from the applied pressure gradient $I$,
and energy dissipation $D$, each normalised by their respective value for laminar flow \rev{(columns of output \texttt{vel\_totEID.dat})}.

\section{Impact  and conclusions}
\label{}

%\textbf{This is the main section of the article and the reviewers weight the description here appropriately}

%Indicate in what way new research questions can be pursued as a result of the software (if any).

%Indicate in what way, and to what extent, the pursuit of existing research questions is improved (if so).

%Indicate in what way the software has changed the daily practice of its users (if so).

%Indicate how widespread the use of the software is within and outside the intended user group.

%Indicate in what way the software is used in commercial settings and/or how it led to the creation of spin-off companies (if so).

%\section{Conclusions}
%\label{}

%Set out the conclusion of this original software publication.

The \OPF  solver aims to provide a fast but 
flexible code, that can
be use for state-of-the art research in the study of 
turbulent flows and transition.  
%As the flow also produces interesting chaotic dynamics in artificially 
%small domains, it may also be used for entry into the world of simulation
%and dynamical systems.
%, and for developments of
%new techniques in research.

Pipe flow is a classical setting for the development of methods for
modelling and analysing dynamical systems, and \OPF has been
used by several groups around the world to make an important 
contribution to developments 
in our understanding of subcritical transition, 
e.g.\ \cite{AvMeRoHo13,ChWiKe14,barkley2015rise,ShHsGo15,Pringle09,WillKer08}.
%IST AUSTRIA, ERLANGEN/BREMEN, 
%GOLDENFELD, JAPAN, LIMSI, BRISTOL, COVENTRY, 
%1000 HITS/MONTH AND RISING.

From these developments have arisen many new opportunities.  
From the theoretical viewpoint, open issues relate to 
comprehension of the role of newly discovered equilibria and periodic
orbits.  Such states are believed to provide a skeleton for the dynamics,
but describing the topology of the state space for turbulence remains 
a challenging and active area.  Pipe flow, and the study of shear
flows in general, draw interest from a range of branches of 
mathematics and theoretical physics, 
e.g.\ pattern formation, control theory,
statistical physics, experimental physics.  
It is an active area of 
cross-fertilisation for the 
development of mathematical and numerical methods.

From a more practical viewpoint, the dynamical
systems approach is being applied
in the modelling of other important flows, e.g.\ flows of
fluids of complex rheology, e.g.\ stress-dependent viscosity, 
particulate flows and
multiphase flows.  The study of `high Reynolds number' flows
is also being influenced via application of dynamical 
systems techniques using LES.

\OPF stands well placed to 
make an increasingly valuable contribution to this effort.
Alongside the application of methods drawn from chaos theory,
extensions to \OPF have just been added for
shear-thinning fluids and LES, for example.
From a research perspective, 
plenty of exciting new developments are in the pipeline.

\section*{Acknowledgements}
\label{}
%Optionally thank people and institutes you need to acknowledge. 
%\input ../Acknowledgements
The author would like to acknowledge John Gibson (\verb+channelflow.org+), 
%the founder of channelflow.org, %for assistance in setting up openpipeflow.org,
%plus 
Predrag Cvitanovi\'c (\verb+chaosbook.org+), Rich Kerswell
and many others for help and inspiration.
The author is also very grateful for financial support from 
EPSRC GR/S76144/01, EP/K03636X/1 and the E.U.~FP7\,PIEF-GA-2008-219-233.
This article was completed at the Kavli Institute for Theoretical Physics,
supported in part under Grant No.~NSF PHY11-25915.



%% References:
\bibliographystyle{elsarticle-num} 
\bibliography{pipes}
\end{document}